\documentclass[aps]{revtex4}
\usepackage{amsfonts}
\usepackage{amsmath}
\usepackage{amssymb}
\usepackage{graphicx}

\begin{document}

\author{A. P\'{e}rez Mart\'{\i}nez}
\email{aurora@icmf.inf.cu} \affiliation{Centro Brasilero de
Pesquisas Fisicas, Instituto de Cosmologia, Relatividade e
Astrof´ýsica (ICRA-BR), Rua Dr. Xavier Sigaud 150, Cep 22290-180,
Urca, Rio de Janeiro, RJ}

\affiliation{Instituto de Cibern\'{e}tica Matem\'{a}tica y F\'{\i}sica (ICIMAF) \\
Calle E esq 15 No. 309 Vedado, Havana, 10400, Cuba}

\author{H. P\'{e}rez Rojas}
\email{hugo@icmf.inf.cu}
\affiliation{Instituto de Cibern\'{e}tica Matem\'{a}tica y F\'{\i}sica (ICIMAF) \\
Calle E esq 15 No. 309 Vedado, Havana, 10400, Cuba}

\author{H. Mosquera Cuesta}
\email{hermanjc@cbpf.br} \affiliation{Instituto de Cosmologia,
Relatividade e Astrof´ýsica (ICRA-BR), Centro Brasileiro de
Pesquisas F´ýsicas Rua Dr. Xavier Sigaud 150, Cep 22290-180, Urca,
Rio de Janeiro, RJ}
\title{Anisotropic Pressures in Very Dense Magnetized Matter }


\begin{abstract}
The problem of anisotropic pressures arising  as a consequence of
the spatial symmetry breaking introduced by an external magnetic
field in quantum systems is discussed. The role of the conservation
of energy and momentum of external fields as well as of systems
providing boundary conditions in quantum statistics is considered.
The vanishing of the average transverse momentum for an
electron-positron system in its Landau ground state is shown, which
means the vanishing of its transverse pressure. The situation for
neutron case and Strange Quark Matter (SQM)  in $\beta$-equilibrium
is also briefly discussed.  Thermodynamical relations in external
fields as well as the form of the stress tensor in a quantum
relativistic medium are also discussed. The ferromagnetic symmetry
breaking is briefly discussed.
\end{abstract}
\maketitle

\section{Introduction}
The idea of local anisotropy in self gravitating systems is frequently
discussed in the literature \cite{Herrera} , although usually it is not
studied how it is generated.  However, the notion of local pressure
anisotropy, that is, the occurrence of unequal principal stresses is a
natural consequence of the spatial anisotropy introduced by external
fields, as is evident to everybody to occur in atmospheric, earth crust
and ocean pressures, where a preferred direction of increasing (radial)
pressure is due to the (approximately) centrally-symmetric gravity
force.  This radial anisotropy is observed to occur, for instance, when
considering a macroscopic sphere in any of these media. In a small
neighborhood of any point inside it, the pressure seems to be
isotropic, but actually this is not so, and it only remains unchanged as
we move on isobar surfaces. But as we move perpendicularly to these
surfaces, it changes due to the momentum added by the external gravity
force field. A very well known anisotropy in pressures is present in
rotating bodies, as planets and stars, since their mass is under the
action of an axially symmetric centrifugal pressure which is added
vectorially to the gravity force, and produces the flattening of these
bodies.

For the case of a gas of electrically charged particles in an
external constant homogeneous magnetic field $B$ (which is field we
will consider along this paper. For variable fields most of our
results are valid for small volumes and/or short intervals of time),
in classical electrodynamics, it is the Lorentz force ${\bf F} =
e{\bf v} \times {\bf B}/c$ the source of an asymmetry in the
pressures parallel and perpendicular to ${\bf B}$ acting on the
particles \cite{Martinez:2003dz}. The Lorentz force stems from the
fact that an electrically charged spinless particle, which in its
interaction with the external magnetic field, moves in a way
equivalent to an effective current, which in turn, generates a
magnetic field opposite to $B$. That is the content of the Lenz law.
By writing $e{\bf v}={\bf j}\Delta V$, where $\Delta V= dx_1 dx_2
dx_3$, calling $f_i = F_i/\Delta V$ as the $i$-th component of the
force density, and substituting ${\bf j}=c \bf{\nabla}\times {\cal
M}$, one has

\begin{equation}
f_i = -(\partial_i {\cal M}_s)B_s + (\partial_s {\cal M}_i)B_s
\label{LFD}
\end{equation}

\noindent multiplying by $\Delta V= dx_1 dx_2 dx_3$ and assuming
$B_s= B\delta_{s3}$ and $\partial{\cal M}_i/\partial x_3=0$
(actually it is also ${\cal M}_i= {\cal M}\delta_{i3}$), only the
first term in (\ref{LFD}) remains as nonzero and one gets back an
expression for the force. For the pressure perpendicular to the
field it results $p_{\perp}= - {\cal M}\cdot \bf{B}$. This is a
classical effect and obviously $p_{\perp}$ must be added to the
usual kinetic isotropic pressure. As in classical electrodynamics
${\cal M}$ is opposite to ${\bf B}$, then ${\cal M}\cdot {\bf B}<0$,
and $p_{\perp}>0$. Its effect is similar to the centrifugal force
mentioned above.  However, the opposite case occurs when ${\cal
M}>0$ (i.e., in the paramagnetic or ferromagnetic cases) being
oriented along ${\bf B}$, which occurs in the quantum case due to
the coupling of the elementary magnetic moments due to spin with the
external field. This quantum effect cannot be derived, obviously,
from the classical Lorentz force.

After our papers \cite{Martinez:2003dz}, \cite{Chaichian:1999gd},
claims were made against the results by some authors, as
\cite{Espinosa:2003pk}, \cite{Khalilov}) starting from classical
results. The contradiction stems from the fact that usually the
classical stress tensor is derived from the Lorentz force
\cite{Jackson}. But in a degenerate quantum gas the dynamics is
described by the Dirac equation, and the spin coupling to the
magnetic field plays a fundamental role (\textit{that is, the
particle does not obey the classical Lorentz force equation}),  and
the use of the classical stress tensor may lead to contradictory
statements. Obviously, if spin is ignored, one might obtain results
in agreement with classical electrodynamics. These are in
correspondence with the classical collapse case discussed in
\cite{Martinez:2003dz}.

In  papers \cite{Martinez:2003dz}, \cite{Chaichian:1999gd} it was
proposed the occurrence of a quantum magnetic collapse based in the
fact that for a relativistic degenerate fermion gas placed in a very
strong external magnetic field, the pressure perpendicular to the
field direction vanish for a field strong enough. The degenerate
fermion gas is composed either of charged particles as electrons and
positrons, or by neutral particles having non-zero magnetic moment,
as it is a gas of neutrons in a background of electrons and protons.
The collapse may occur since for such a gas: 1) its pressure is
exerted anisotropically, having a smaller value perpendicular than
along the magnetic field and 2) there are critical values of the
magnetic field strength for which the magnetic response of the gas
through its magnetization is such that it produces the vanishing of
the equatorial pressure of the system, and the outcome would be a
transverse collapse of the star.  Several authors (i.e.
\cite{Kohri:2001ah}, \cite{Mosquera})  have referred to our
proposal.

We must emphasize that such effect means that for any body, the
dimensions transverse to $B$ decrease with increasing $B$. It is in
correspondence with the fact that in the quantum regime a magnetic
field leads to a characteristic length $\sqrt{\hbar c/eB}$ which
gives a measure of the wave function spread perpendicular to $B$,
and it decreases with increasing $B$. The \textit{quantum area}
$\hbar c/eB$ has the property that the magnetic flux through it
leads to a quantum flux $B(\frac{\hbar c}{eB})=\frac{\hbar c}{e}$
and for increasing $B$ such area decreases.  There is a shrinking
orthogonal to $B$ which has several consequences in microscopic
quantum systems. For instance, in quantum vacuum, for any body
placed on a strong magnetic field it is produced a shrinking
perpendicular to the field and a stretching along it \cite{Perez
Rojas:2006dq}, and recently it has been shown \cite{Shabad:2007zu}
that the Coulomb potential of a charge placed on a strong magnetic
field follows an anisotropic law, decreasing away from the charge
slower along $B$ and faster across it. For very large magnetic
fields the potential is confined to a thin string passing though the
charge parallel to $B$.


Also, a model based on the dynamic description of a local volume of
a magnetized self--gravitating Fermi gas starting from a very dense
Fermi gas, it has been also shown recently the occurrence of a
quantum magnetic collapse along a strong external magnetic field
\cite{Ulacia Rey:2007kc}.

We would like  in the present paper to discuss some points about the
role of the energy-momentum tensor on quantum relativistic matter in
a magnetic field. We will refer mainly to the electron-positron gas
in a magnetic field, as we did in \cite{Chaichian:1999gd}, but will
refer also to the neutron gas, as a model for neutron stars and we
also will comment the magnetized quark matter in the framework of
the phenomenological MIT Bag Model to describe Strange Stars
\cite{Perez Martinez:2005av}-\cite{Felipe:2007vb}.

In the electron-positron case, we show explicitly the vanishing of
the squared transverse momentum average, for the particle in the
Landau ground state, leading to a vanishing pressure. We are
thinking of a model of a white dwarf (or neutron star), in which it
is assumed that the dominant pressure is due to the electron
(neutron) gas, which can be described as a degenerate quantum gas.
For the white dwarf we assume, thus, that there is a nuclei
background which compensates the electric charge, but whose
thermodynamics is described classically and leads to quantities
negligibly small as compared with those of the electron gas. This
means that we use the results of relativistic quantum statistics in
equilibrium, applied to the study of the properties of an magnetized
fermion gas in an external field. We start, thus, from a quantized
theory in presence of an external field, as  is the problem of
electrons in a magnetic field (or either in an atom), where the
quanta of the fields involved are the electron-positron and photon
fields.

The external fields are, obviously, not considered of quantum
nature, and its physical consequences are manifested through their
interactions with the quanta, expressed by appropriate terms in the
Lagrangian.

We discuss also the thermodynamics for the axially symmetric systems
in a magnetic field, having anisotropic properties, and to the fact
that material bodies deform to adopt the appropriate shapes to
maintain equilibrium, although in some cases the system may become
unstable \cite{Ulacia Rey:2007kc} and collapse. Finally we also
discuss briefly the form of the electromagnetic stress tensor in
relativistic matter, in particular for a self-magnetized system.

\section{Quantum mechanics and quantum statistics in external fields}

External fields and boundaries have a close analogy. It is usual
in the most elementary problems of statistical mechanics, as that of
an ideal gas, to assume that there is momentum conservation in
molecular collisions. However, the gas is confined to a vessel
with which the molecules exchange momentum (and energy), and total
conservation of momentum and energy cannot hold unless the vessel
is included, as a thermostat. The conservation of energy and
momentum is, nevertheless, assumed to take place in the gas
\textit{on the average}. Thus, one speak about internal energy and
pressures inside the gas, under the previous assumptions. Usually,
the energy and momentum of the walls of the vessel do not enter
into play in the thermodynamic description of the gas. However,
the role of the vessel appears in fixing the external parameter
$V$.

Another point is the inclusion of the classical energy of the
electromagnetic field in quantum mechanical calculations. If we
calculate the electron energy in the atom, we should \textit{not}
add to the usual Dirac electron energy, the term $T_{44}$ coming
from the stress tensor produced by the external field of the
nucleus. This field is $E= Z e^2/r^2$, and if its square divided by
$8\pi$ is integrated from some small $\epsilon$ to infinity, we
would get an expression of form $U=Z^2 e^2/2\epsilon$, which
diverges as $\epsilon \to 0$. This would lead to  the expression of
\textit{classical } electromagnetic self-energy of the nucleus.
There is no reason for adding this energy to the Dirac electron
energy eigenvalue. This is nonsense according to standard quantum
theory. In a similar way, there is no reason for adding such energy
to the quantum statistical average in a problem of many electrons.
However, in dealing with the total system (electrons $+$ nuclei),
one must consider the problem of self-energy both of electrons and
nuclei from the \textit{quantum} point of view. We must consider
also conservation of energy and momentum of the whole system.

For the case of electrons moving under the action of an external
magnetic field, the problem is similar. The dynamics of the electron
is described by its energy eigenvalues and the classical energy of
the external magnetic field is a quantity that ultimately
corresponds to the energy of the source of the external field (for
instance, the energy of the current creating it). As we will see
below, all the dynamics of the electrons come from the interaction
term of the electron-positron field with the total electromagnetic
field (the external field plus the radiation field $A_{\mu}^{ext}+
a_{\mu}$) in the Lagrangian, and later in the equations of motion,
and there is no any reason for adding to the electron energy
eigenvalues (and/or to the statistical energy or momentum average)
the classical energy and/or momentum of the external field.

Obviously, in considering the conservation laws applied to the total
system, it is necessary to take into account the energy and momentum
of the sources of the field. These are usually confined to a small
region of the total system, and frequently it is possible to
describe them classically. In most of the volume of the system,
however, the electrons are to be considered as being under the
action of an external field, and for them, for instance, the
momentum transverse to the field is not conserved.

The thermodynamics of an electron gas in an external field is, in
consequence, described also by its interaction term with the
electron-positron field in the Lagrangian. For a charged particle in
presence of an external magnetic field $B$ (parallel to $x_3$)  the
momentum component along the field $p_3$ is a quantum number, which
together with the Landau quantum number $n(=0,1,2..)$ (which do not
correspond to momentum eigenvalues), and the spin component along
$B$ (the eigenvalues $\pm 1$ of $\sigma_3$, characterizes the set of
quantum states. The latter are degenerate with regard to a third
quantum number, the orbit's center coordinate $x_0=p_i/eB$, where
$i=1$ or $2$ depending on the gauge choice for the four-potential
from which the external field $B$ is derived. For a neutral particle
with nonzero magnetic moment $q/m$, the three components of momentum
are observable, but they do not enter symmetrically in the energy
spectrum. The component $p_3$ behaves as in the zero field case, but
the transverse momentum $p_{\perp}=\sqrt{p_1^2+p_2^2}$ enters in a
different way (see below), and for a magnetic dipole moment parallel
(antiparallel) to $B$ the contribution is decreased (increased) as
$B$ grows. From all this, we conclude that an anisotropy in the
quantum dynamics must be reflected in an anisotropy in the
thermodynamical properties of the quantum gas in a magnetic field.
This leads to the necessary formulation of the statistical physics
of the problem taking into account the arising of anisotropic
quantities.

The quantum statistics of the problem is made by taking the average
of the field operators by using the density matrix operator, which
is equivalent to a quantization at finite temperature and nonzero
density. The extremely degenerate case corresponds to the zero
temperature case. We start from the Lagrangian describing the
interaction of the electron-positron and photon fields in presence
of an external magnetic field. A background of positive charge, due
to the nuclei, is assumed to exist to compensate the electron
charge. Its Lagrangian (which usually gives a negligible
contribution), as well as the Lagrangian of the subsystem generating
the external magnetic field, are not included in the present
approximation, which considers only the interaction of the electron
gas with a given external magnetic field (assumed as constant and
uniform). However, in some cases (for instance, if we consider
vector bosons with non-zero magnetic moment \cite{ChJP}),
\cite{IJMPD}, the system may generate self-consistently its magnetic
field, and it is not necessary to consider external sources for it.
In the present electron case, we have,

\begin{equation}
{\cal{L}} = \bar\psi[\gamma _{\mu}(\partial_{\mu}-ieA_{\mu})-m]\psi
+\frac{1}{16 \pi}\textsl{F}_{\mu\nu}^{2} \label{1}
\end{equation}

Here $A_{\mu}=A_{\mu}^{ext}+ a_{\mu}$, where
$A_{\mu}^{ext}=Bx_{1}\delta_{2\mu}$ is the external vector
potential in some fixed gauge, and $a_{\mu}=a_{\mu}(x)$ is the
radiation field (in the case of a relativistic electron in the
atom, it is usually taken
$A_{\mu}^{ext}=-\frac{Ze^{2}}{r}\delta_{0\mu}$, in which it has
been fixed also a gauge for the external field). In (\ref{1}) we
exclude terms linear in the fields, whose quantum average is zero.
Thus, the pure field term is to be replaced by the sum of the
external field tensor squared plus the radiation field tensor
squared, i.e.
$\textsl{F}_{\mu\nu}^{2}\rightarrow\textsl{F}_{\mu\nu}^{ext2}+
\textsl{f}_{\mu\nu}^{2}$. The equations of motion are

\begin{equation}
[\gamma_{\mu}(\partial_{\mu}-ieA_{\mu})-m]\psi=0\; , \hspace{0.5cm}
\frac{\partial\textsl{f}_{\mu\nu}}{\partial x_\nu}=\bar \psi
\gamma_{\mu} \psi \label{2} \; ,
\end{equation}

\noindent where we see that $\textsl{F}_{\mu\nu}^{ext}$ obviously do
not enter in the equations of motion: we did not included in our
problem the external sources generating physically the external
field. We may study, however, the problem of the electron-positron
gas in the constant and homogenous external magnetic field
consistently, since the interaction of matter with the external
field is properly included. Even more, the equations of motion
(\ref{2}) do not contain the contribution from the pure external
field term $\frac{1}{16 \pi}\textsl{F}_{\mu\nu}^{ext
2}=\frac{1}{8\pi}B^2$ (recall that the external field $\textbf{B}$
remains as only magnetic in all frames of reference moving parallel
to it, but the quantity $\frac{1}{8\pi}\textsl{F}_{\mu\nu}^{ext 2}$
is a Lorentz invariant), since such term is independent of the
fields to be quantized $\bar\psi, \psi, a_\mu$ and of the
coordinates $x_\mu $. Physically, this means that the particles do
not feel the momentum (or pressure) and energy generated by that
term, but only the quantities coming from the generalized
four-momentum $p_\mu + eA_\mu $, or equivalently, from the
interaction with the field $ie \bar \psi \gamma_{\mu}A_{\mu}\psi$.

\section{Average transverse momentum and pressure of the electron gas }

 The vanishing of the transverse pressure in the Landau ground
state can be seen also from quantum mechanical considerations. The
electrons (positrons) in an external magnetic field $B$ have energy
eigenvalues $E_{n,p_3,\pm 1}=\sqrt{m^2+p_3^2+2eB(n+1/2)\pm eB}$, and
its quantum states are described by the spinor wavefunctions
$\Phi_{n}^{\pm}(x)$ \cite{Johnson}

\begin{equation}
\Phi_{n,p_2, p_3, 1}^{\pm}(x)=\left[\frac{E(p_3,n)\pm m}{2
E(p_3,n)}\right]^{1/2}\frac{e^{ip_2 x_2 +ip_3 x_3}}{2 \pi}
\left[\begin{array}{c}
  \varphi_{n-1}(\xi) \\
  0 \\
  \frac{p_3 \varphi_{n-1}(\xi)}{m \pm E(p_3,n)} \\
  \frac{i(2n eB)^{1/2} \varphi_{n}(\xi)}{m \pm E(p_3,n)}
\end{array}\right] \label{J1}
\end{equation}

and
\begin{equation}
\Phi_{n,p_2,p_3 -1}^{\pm}(x)=\left[\frac{E(p_3,n)\pm m}{2
E(p_3,n)}\right]^{1/2}\frac{e^{ip_2 x_2 +ip_3 x_3}}{2 \pi}
\left[\begin{array}{c}
  0 \\
 \varphi_{n}(\xi) \\
 -\frac{i(2n eB)^{1/2} \varphi_{n-1}(\xi)}{m \pm E(p_3,n)}\\
 - \frac{p_3 \varphi_{n}(\xi)}{m \pm E(p_3,n)} \\
\end{array}\right] \label{J2}
\end{equation}

\noindent where the superscript $\pm$ refers respectively to
positive and negative energy solutions, and (\ref{J1}),(\ref{J2})
correspond  to the eigenvalues $\pm 1$ of $\sigma_3$, respectively
parallel and antiparallel to $B$. For $n=0$, only the second (spin
down) term contributes. It contains as a factor a term proportional
to the Hermite function of zero order $(eB)^{1/4}A_0
e^{-\xi^2/2}e^{i p_2 x_2 +i p_3 x_3}$ of argument $\xi =
(eB)^{1/2}(x+x_0)$, where $x_0 = p_2/eB$ is the coordinate of the
orbit's center. The energy eigenvalues are degenerate with regard to
$x_0$. Expressions (\ref{J1})-(\ref{J2}) are obviously not
eigenfunctions of the transverse momentum operators
$i\partial/\partial_{1,2}$. The transverse momentum eigenfunctions
would be given by an infinite series in terms of
(\ref{J1}),(\ref{J2}) and one can easily check that for the system
of non-interacting electrons, if confined to the Landau ground state
$n=0$, the expectation value of its effective momentum operator
perpendicular to the field is zero.  This is especially interesting
because as the magnetic field increases the number of occupied
Landau states tends to decrease up to the limit $n=0$. Since
$\partial\varphi_n (\xi)/ \partial\xi =
\sqrt{n/2}\varphi_{n-1}(\xi)-\sqrt{(n+1)/2}\varphi_{n+1} (\xi)$, and
by defining the squared transverse momentum operator as the quantity
$[-\partial^2/\partial x_{1}^2+ \sigma_3 eB/2]$, where we have added
the spin contribution $\sigma_3 eB/2$ to the second derivative with
regard to $x_1$  (Obviously, the spin term along the $3$-axis we
assume as contributing to the momentum perpendicular  to $B$. We did
not included the derivative with regard to $x_2$ since $p_2$ leads
to the coordinate $x_0$ of the center of the orbit, and the energy
eigenvalue is degenerate with regard to it) one can write the
average effective transverse momentum squared $p_{\perp}^2$ when the
system is in the Landau ground state $n=0$, $\sigma=-1$, as

\begin{equation}
\int \bar\Phi_{0,p_{2,3}}^{\pm}(x)[-\partial^2/\partial
x_{1}^2-eB/2]\Phi_{0,p_{2,3}}^{+}(x) d x_1=0. \label{K}
\end{equation}

\noindent  The transverse effective momentum in the ground state is
thus $p_{\perp}=\sqrt{E_0^2-p_3^2-m^2}=0$  and the electron gas
behaves as a one dimensional (parallel to $B$) gas and does not
exert any pressure perpendicular to $B$.  This effect becomes
important for densities of order $\lambda^{-3}_C\sim 10^{30}$
cm$^{-3}$ ($\lambda_c$ is the Compton wavelength) and fields near
the critical value $B\sim B_c$, where $B_c \sim m^2 c^3/e\hbar \sim
10^{13}$ Gauss. This implies a stretching of the body along the
$B$-field direction and a shrinking perpendicular to it. (The effect
of the anomalous magnetic moment of the electron may appear if we
consider radiative corrections, i.e., calculations beyond the
present tree level).

If one takes into consideration the contribution from quantum
vacuum, as it must be,  \cite{Perez Rojas:2006dq}, one finds that
actually, the pressure resulting in the electron case for Landau
quantum number $n=0$, must be increased in a \textit{negative
quantity}, since the transverse pressure of quantum vacuum is
negative, which enhances the shrinking effect of the gas.

\section{The neutron gas case}

For the neutron gas, however, the anomalous magnetic moment plays an
important role already at the tree level. Due to it, the anisotropy
in pressures is similar to the one in the electron-positron gas,
leading to a stretching of the body along the magnetic field. As
pointed out in \cite{Martinez:2003dz}, the neutron eigenvalues in a
magnetic field

\begin{equation}
E_n(p,B,\eta )=\sqrt{p_3^2+(\sqrt{p_{\perp}^2+m_n^2}+\eta qB)^2},
\end{equation}

\noindent where $\eta=\pm$ and $q=1.91M_n$, where $M_n$ is the
nuclear magneton. A degenerate neutron gas exhibit the relativistic
behavior analog to Pauli paramagnetism. For adequate values of the
density and magnetic field, the amount of neutrons with magnetic
moment up ($\eta=-1$) is largely greater than those with magnetic
moment down ($\eta=+1$). This results from the discussion of the
densities of neutrons $N_n^{\pm}$ with magnetic dipoles up and down
($\eta=\mp 1$), by starting from the Fermi distribution for neutrons
in the degenerate case, $n_n(E_n)=\theta (\mu_n -E_n(B, \eta=\mp
1))$. The vanishing of the argument of the $\theta(x)$ functions
define the two Fermi surfaces for $\eta=\mp 1$, and the boundaries
of the integral for $N_n$ below, which is taken in cylindrical
coordinates in momentum space

\begin{eqnarray}
N_{n}^{\pm}& =& \frac{1}{4 \pi^2}\sum_{\eta=1,-1}
 \int_{0}^{\sqrt{(\mu-\eta qB)^2-m_n^2}}p_{\perp}dp_{\perp}\int_{-p_{3F}}^{p_{3F}} dp_3\theta \left(\mu -
E_n (p_{\perp},dp_3, B, \eta)\right)  . \label{ONT1}
\end{eqnarray}

\noindent Where $p_{3F}=\sqrt{\mu^2-(\sqrt{p_{\perp}^2+m_n^2}+\eta
qB)}$. We observe that for $\mu-qB \to m_n$ the upper limit of
integration in (\ref{ONT1}) tends to $0$, which means that $N^{-}$
also tends to zero. We shall consider in what follows $B$ large
enough to satisfy approximately these conditions, so that the term
$N^{+}$ has the main contribution to $N$. The Fermi surface for
$\eta=-1$ is defined by $\mu_n -E_n(B, \eta=-1)=0$. The effective
Fermi transverse momentum squared is

\begin{equation}
p_{F\perp eff}^2 =\mu_n^2-p_{F3}^2-m_n^2 =\left(\sqrt{ p_{F\perp}^2
+ m_n^2}-qB\right)^2-m_n^2.\label{pF}
\end{equation}
\noindent The second term of (\ref{pF}) decreases with increasing
$B$, and  may even vanish for some critical fields. For instance, if
$B \ll B_{cn}$, where $B_{cn}= m_n/q \sim 10^{20}$ G, then $q^2 B^2
\ll 2 q B m_n$. In such case the vanishing of $p_{F\perp eff}$ is
guaranteed if $p_{F\perp} \sim \sqrt{2q B m_n}$. Thus, the neutron
gas statistical behavior under these physical conditions is
equivalent to a one-dimensional gas, with effective spectrum
$E_n(p,B)=\sqrt{p_3^2+m_n^2}$. The transverse pressure vanishes. For
$p_{F\perp}\sim 10^{-1.5} m_n $, $qB/m_n \sim 10^{-3}$, which means
fields of order $B \sim 10^{17}$G.

\section{Magnetized Strange Quark Matter}

Strange quark matter (SQM), which means to have quarks $u$ $d$ and
$s$ and electrons, could be studied using the MIT bag model
~\cite{Perez Martinez:2005av}. In that model, confinement is
guaranteed by the bag and quarks are considered as a Fermi gas of
noninteracting particles. Under these assumptions, it is possible to
study the thermodynamical properties of a quark gas in a strong
magnetic field. In our study  the anomalous magnetic moment (AMM) is
included, which means that Pauli paramagnetism is taken into account
besides Landau diamagnetism, given by the presence of Landau levels.
In that case the  energy spectrum  lose its degeneracy. The
thermodynamical quantities depends on two sums: one over Landau
levels and other over the two orientations of the spin parallel or
antiparallel to the magnetic field. As the particles (quarks)  have
positive or negative AMM, they have different preferences in the
spin orientation with respect to the magnetic field. This has
important consequences in the EOS of the system. The most relevant
comes from the energy ground state, which depends on the strength of
the magnetic field and it could be zero (cf. Eq.~(\ref{energy1})).

\begin{align}
E_{i,n}^{\eta}=\sqrt{p_3^2 + m_i^2 \left( \sqrt{
\frac{B}{B^c_i}(2n+1-\eta) + 1} - \eta Q_{i}B
\right)^2},\label{energy}
\end{align}

\noindent
with $B^{c}_i=\frac{m_{i}^2}{|e_i|}$ $i=(e,u,d,s)$, $e_i$
and $m_{i}$ denote the charges and the masses of the particles,
respectively. The quantities $Q_{i}$ are the corresponding AMM of
the particles,
\begin{align}
Q_e&=0.00116\mu_B\,,\quad Q_u=1.85\mu_N\,, \nonumber\\
Q_d&=-0.97\mu_{N}\,,\quad Q_s=-0.58\mu_{N}\,,
\end{align}
where
\begin{align}
\mu_{B}&=\frac{e}{2m_e} \simeq 5.79 \times 10^{-15}\, {\rm MeV/G}\,,\nonumber\\
\mu_{N}&=\frac{e}{2m_p} \simeq 3.15\times 10^{-18}\, {\rm MeV/G}.
\end{align}

\noindent
$m_{u}=m_{d}=5$~MeV and $m_{s}=150$~MeV for the light
quark masses. The magnitudes of the so-called critical fields
$B^{c}_i$ (when particle's cyclotron energy is comparable to its
rest mass) are $B^{c}_e= 4.4\times 10^{13}$~G, $B^{c}_u= 6.3\times
10^{16}$~G, $B^{c}_d=1.3\times 10^{16}$~G and $B^{c}_{s}=1.1\times
10^{19}$~G.

It can be seen from the spectra (\ref{energy}) that, besides of the
quantization of their orbits in the plane perpendicular to the
magnetic field, charged particles with AMM undergo the splitting of
the energy levels with the corresponding disappearance of the
spectrum degeneracy. For the non-anomalous case, $Q_{i}=0$, the
ground state energy is independent of the magnetic field strength
and the magnetic field only quantizes the kinetic energy
perpendicular to the field similar to that of the electron case. In
this situation, the energy is degenerate for Landau levels higher
than zero. States with spin parallel or antiparallel to the magnetic
orientations ($\eta=\pm 1$) have the same energy. However, the
anomalous case, $Q_i\neq 0$, removes this degeneracy. In the latter
case, the rest energy of the particles depends on the magnetic field
strength. The ground state energy is
\begin{align}
E_{i,0} = m_i  \left(1-y_i B\right) .\label{energy1}
\end{align}

The above equation leads to the appearance of a threshold value for
the magnetic field at which the effective mass vanishes, $m_i \sim
|Q_i| B$. The thresholds of the field, $B^s_i=m_i/Q_i$,  for all the
constituents of the SQM are given by
\begin{align}
B^{s}_e&=7.6\times 10^{16}\,\text{G},\quad B^{s}_u=8.6\times 10^{17}\,\text{G},\nonumber\\
B^{s}_d&=1.6\times 10^{18}\,\text{G}, \quad B^{s}_s=8.2\times
10^{19} \,\text{G},
\end{align}
that are smaller than the ones obtained when the classical AMM
contribution is considered~\cite{Chakrabarty:1996te}.

The expression (\ref{energy1}) suggests that the energy of the
particles becomes smaller than that of the antiparticles, which
 which might be manifested in the creation of pairs.
As spontaneous pair creation in a magnetic field seems to be
forbidden, for individual particles, the correct meaning of this
"critical" field is that it corresponds to an upper bound.

In the SQM scenario all the constituents interact with the magnetic
field and are obliged to satisfy the equilibrium conditions. Under
such constraints, it turns out that the dominant threshold field
comes from $u$ quarks, thus leading to the upper bound $B \lesssim
8.6 \times 10^{17}$~G. This result has an important astrophysical
consequence, since the bound for SQM can be also extrapolated to the
SQS scenario. If  SQS exist, the maximum magnetic field strength
that they could support  would be around the above bound, i.e.
$10^{18}$~G.

 The density of particles, defined as $N=\sum_i N_i$ with
$N_i=\frac{\partial\Omega_{i}}{\partial\mu_i}$ gives

\begin{equation}
N_{i}^{\pm} =\frac{1}{4 \pi^2}\sum_n^{nmax}\sum_{\eta=1,-1} \int
_0^{\sqrt{(\mu-\eta
Q_iB)^2-eB(2n+1-\eta)-m_i^2}}\int_{-p_{3F}}^{p_{3F}}
 dp_3\theta \left(\mu_i -
E_i (p_{\perp},dp_3, B, \eta)\right)  . \label{ONT2}
\end{equation}

\noindent
with

\begin{align}
p_{F\,\perp,i}^2=\mu_i^2-p_{F\,3}^2-m_i^2=\left(\sqrt{e_iB(2n+1-\eta)
+ m_i^2} -\eta Q_{i}B\right)^2 -m_i^2,\label{e2}
\end{align}

The presence of AMM produces that  $p_{F\perp\,eff\,i}$ decreases
with the increasing of magnetic field. For $n=0$ we have that

$$p_{\perp}=BQ_i|(BQ_i-2m_i)|$$

\noindent which would be zero for $B=2m_i/Q_i$.

Nevertheless, for the SQM in $\beta$-equilibrium never is reached
the condition $p_{\perp}=0$. It is due to the restriction imposing
by $\beta$-equilibrium that makes that appear an upper bound for the
magnetic field that is one half of this one that does $p_{\perp}=0$.
This condition is reached because the chemical potential of
electrons involved in SQM becomes negative which is  lacking of
physical sense. If AMM is not take into account we obtain the
condition $p_{\perp}=0$ when $n=0$ for fields around $10^{19}$~G
\cite{Perez Martinez:2005av}.

From the quantum statistical point of view the lowest energy states
with AMM contain important physical consequences: for particles with
mass $m_i$ and anomalous magnetic moment $Q_i$, the magnetic field
has a critical value given by the expression $B_i^s \sim m_i/|Q_i|$.
For SQM in $\beta$-equilibrium it turns out that the dominant
threshold field comes from $u$ quarks, thus leading to the upper
bound $B \lesssim 8.6 \times 10^{17}$~G \cite{Felipe:2007vb}.


\section{Thermodynamics for Anisotropic systems}

In this section we shall assume that we are in a reference frame in
which the center of mass of the system we are considering is at
rest. Quantum statistics is currently formulated for isotropic
systems. In presence of an external electromagnetic field, the
entropy $S$ and the energy $U$, which are basic scalars, are
dependent on the tensor field ${\cal F_{\mu\nu}}$.  The external
field breaks the spatial isotropy and this is reflected in the
quantum mechanical observables. In standard statistical mechanics,
in which the system is assumed as isotropic, the usual
interpretation of thermodynamic quantities is through the relation
$\Omega=-PV$, where $-P=\Omega_V=-\beta^{-1}\ln {\cal Z}$ (we are
naming $\Omega_V=\Omega/V$). Here we assume that ${\cal Z}$ is built
from a density matrix $\rho=e^{-\beta\int d^3 x({\cal H}-\mu N)}$
where the integral is taken over all space. There is one external
parameter $V$, and one conjugate variable $P$. Hidden in this
isotropic formulation is the idea of the elementary work $\delta W =
f_i dx_i$ done by a generalized force $f_i$ in the direction of the
elementary displacement $dx_i$. This generalized force comes from
the product of the isotropic pressure tensor $P_{ij}=p \delta_{ij}$
by the surface element $dS_j$, e.g. $f_i=P_{ij}d\sigma_j=p dS_i
dx_i=pdV$

In the anisotropic case, $P_{ij}$ even if having a diagonal form,
have in general different eigenvalues. The elementary work comes
from the product of the force $f_i= P_{ij}d\sigma_j$ by the
elementary displacement $dx_i$. One gets $\delta W = f_i dx_i
=P_{ij}d\sigma_j dx_i$. Here the product of the differentials
behaves formally as a "volume pseudo-tensor" $dV_{ji}=d\sigma_j
dx_i$ (This is restricted to three dimensional space. Our
considerations  strictly apply in the frame where the body is at
rest). Actually, the entropy density and the particle density may be
also anisotropic and given by pseudo-tensors $s_{ij}$, $n_{ij}$, so
that $dS=s_{ij}dV_{ji}$, $dN=n_{ij}dV_{ji}$. The differential of the
energy can be written as always, by the sum of the elementary heat
$\delta Q=T dS$ plus the elementary work done by the system $\delta
W=-P_{ij}dS_j dx_i$

\begin{equation}
dU=T dS+\mu dN -P_{ij}dV_{ji}. \label{preani}
\end{equation}

\noindent From (\ref{preani}), one can see that $dU$ may have a
different expression for elementary changes in the volume of the
system taken along different directions in space.

Here arises a very important point of procedure. For isotropic
systems of volume $V$, the thermodynamic potential $\Omega=-PV$ is
the equation of state arising from the evaluation of $\Omega$ in
terms of the Hamiltonian operator, temperature and chemical
potential. There is only one pressure, since the energy-momentum
tensor is spatially isotropic ${\cal T}_{\mu \nu }^i=P\delta _{\mu
\nu }-(P+u)\delta _{4\mu }\delta _{\nu 4}$. However, once we have
anisotropic pressures, such relation is not enough.

When an external magnetic field is present, one can write in terms
of the thermodynamic potential $\Omega_V$ \cite{Shabad},

\begin{equation}
{\cal T}_{\mu \nu }=(T\partial \Omega_V /\partial T+\mu \partial
\Omega_V /\partial \mu )\delta _{4\mu }\delta _{\nu 4}+ 4F_{\mu
\rho }F_{\nu \rho }\partial \Omega_V /\partial F^2-\delta _{\mu
\nu }\Omega_V ,  \label{ten}
\end{equation}

This tensor reduces to the isotropic expression ${\cal T}_{\mu \nu
}^i$ for $F_{\mu \rho }=0$.

\section{Some thermodynamical relations}

For simplicity, we shall assume that shearing stresses are absent.
From the thermodynamical point of view, the new element is that
there are two kind of external parameters, the volume and the
external field, which can be defined either by the electromagnetic
field tensor ${\cal F}_{\mu\nu}$ or by the three dimensional vectors
$\textbf{E}$, $\textbf{B}$, and as conjugate parameters, the
pressures and the electric and magnetic moments $P$, ${\cal M}$. For
instance, in the case of a constant homogeneous magnetic field one
has $\Omega_V=f(\mu,\beta,B)$, and from the relation $U= TS+\mu N +
\Omega$ \cite{Landau} one can write in terms of the extensive
variables variables $S$,$N$,$V_{ij}$, and the intensive variable
$B$,

\begin{equation}
dU=T dS+\mu dN +(\partial\Omega/\partial V_{ij})dV_{ij} +
(\partial \Omega/\partial B)dB. \label{difener}
\end{equation}
\noindent
where $\partial \Omega/\partial B=V\partial
\Omega_V/\partial B=-V{\cal M}$, ${\cal M}$ is the magnetization and
$V {\cal M}$ the magnetic moment. We name again
$\partial\Omega/\partial V=-P$ "the pressure", although care must be
exercised here. The interpretation of this term as the pressure is
free from ambiguities in the isotropic case, where the spatial
diagonal components of the energy-momentum tensor are equal. In
presence of an external field, this is not the situation. Let us
consider the elementary work done for an increase in the volume
$dV_{\perp}$ in the direction orthogonal to $B$. We have

\begin{equation}
\delta W_{\perp} = -(p-{\cal M}B)dV_{\perp}.
\end{equation}

However, along the field, it is $\delta W_{||} =-p dV_{||}$. We have
different results depending on the direction in space. We will
concentrate in the pressures exerted by the electron gas. We obtain
from (\ref{preani}) or (\ref{ten}) different expressions for the
pressure for directions parallel and perpendicular to the magnetic
field,

\begin{equation}
P_3=-\Omega_V ,\hspace{1cm}P_{\perp }=-\Omega_V -B{\cal
M}.\label{tpr}
\end{equation}

The fourth component of the tensor is the expression for the
energy density

\begin{equation}
u=Ts+\mu \rho-\Omega_V, \label{uener}
\end{equation}

where $s=-\partial \Omega_V/\partial T$,$\rho = - \partial \Omega_V
/
\partial \mu$ are respectively the entropy and  particle number
densities.

Concerning the thermodynamical properties of the magnetized gases,
from the definition of ${\cal M}(B)$ we get $\Omega =-\int {\cal
M}dB-P_0$, $P_0$ being the pure mechanical pressure, i.e. the part
of the pressure term independent of the field. One can write

\begin{equation}
dP_0=dP_3- {\cal M}dB. \label{difpremech}
\end{equation}

From (\ref{uener}) and (\ref{difener}) we can write for the
differential of internal energy $U$ when the volume $V$ changes
along $B$, i.e., along the $3$-rd axis as $dV=S_{\perp}dx_3$, where
($S_{\perp}$ is the basis of the volume $V$),

\begin{equation}
dU=T dS+\mu dN - P_3dV - \textbf{M}\cdot d\textbf{B}.
\label{difener1}
\end{equation}

Thus, if $dS=dN=0$ and $B$ is kept constant, the change in internal
energy is the work $\delta W_3=-P_3 dV=-F_3dx_3$, where $F_3=P_3
dS_{\perp}$ is the generalized force along $x_3$. For a displacement
perpendicular to the field one expects that although $B$ and ${\cal
M}$ be kept constant, $\textbf{M}$ must change. To account for it,
we may define another thermodynamic function by means of a Legendre
transformation. By adding $d(\textbf{M}\cdot \textbf{B})$ to
(\ref{difener1}) one gets the differential of the "extended energy"
$U_0$ as

\begin{eqnarray}
dU_0&=&dU+d(\textbf{M}\cdot \textbf{B})=T dS+\mu dN - P_3dV +
\textbf{B}\cdot d\textbf{M}\\ \nonumber &=&T dS+\mu dN - P_{\perp}dV
\; . \label{difener2}
\end{eqnarray}

Also if $dS=dN=dB=0$, the last term in (\ref{difener2}) is the work
done when $dV=S_3 dx_{\perp}$, where $S_3$ is one section of $V$
containing the $3-$rd axis. In other words $dU_0=\delta W_{\perp}=
-P_{\perp}dV=-F_{\perp}dx_{\perp}$, where $F_{\perp}=P_{\perp}S_3$
is the generalized force perpendicular to $B$. We observe that
(\ref{difener2}) has the usual form of the differential of the
internal energy in the zero field case written in terms of the
$P_{\perp}$ pressure. Note that for $P_3 \gg {\cal M}B$, the
$P_{\perp} \sim P_3$ and $U_0$ corresponds to the usual expression
for the internal energy in absence of the field, whereas
$U=U_0-\textbf{M}\cdot \textbf{B} $ expresses the internal energy
after including the interaction of the system, as a dipole
$\textbf{M}$, with the external field $\textbf{B}$.

Finally, by adding $d(P_3 V)$ to $dU$ one has for the differential of
the enthalpy $H=U+P_3V$

\begin{equation}
dH=T dS+\mu dN - V (dP_3 - {\cal M}dB). \label{difenthalp}
\end{equation}

Thus, if $N$ is constant and the pure mechanical pressure is
constant, $dP_0 =0$, $dH$ is equal to the heat absorbed by the
system.

\section{Deformations of bodies due to anisotropy}

We must discuss the physical consequences of the anisotropy in
pressures. Thinking about a star as a model, at any point the gas
pressure is counterbalanced by the pressure due to the gravitational
attraction of the star mass. Let us name this pressure $P_g$. The
equilibrium of the pressures (\ref{tpr}) with $P_g$ leads to

\begin{equation}
P_3=P_g,\hspace{1cm}P_{\perp }=P'_g. \label{equil}
\end{equation}

\noindent The second equation is to be understood as meaning that
$P_3=P'_g + B{\cal M}$. This problem bears some parallelism with the
rotating body in a gravitational field: in that case the effective
gravitational field perpendicular to the axis of rotation is
decreased by the centrifugal force. Here the situation is the
opposite, as the gravitational pressure perpendicular to $B$ is
increased in the amount $B{\cal M}$. The outcome is a prolate isobar
surface, or in other words, the body  stretches along $B$ and
shrinks perpendicular to it, like a cigar. Also, from (\ref{tpr})
and according to \cite{Chaichian:1999gd}, for all the electrons in
the Landau ground state $n=0$ (which does not contradict Pauli's
Principle, since there is a degeneracy with regard to an extra
quantum number, the orbit's center), $-\Omega_V =B{\cal M}$ and
$p_{\perp }=0$ which suggests the occurrence of a collapse of the
system perpendicular to ${\bf B}$.

\section{The electromagnetic stress tensor in relativistic matter and self-magnetization}
 By writing ${\cal M}=(H-B)/4\pi $, one may get formally $\Omega
=-\frac{1}{8\pi }B^2+\frac{ 1}{4\pi }\int HdB-p_0$. We recall that
as $\Omega \equiv F-G$, the last expression is consistent with what
would be obtained in the classical non-relativistic case
\cite{Landau}, where $F=F_0+\int HdB/4\pi $ is the Helmholtz free
energy and $G=F+\int {\cal M}dB+p_0$ $=G_0+B^2/8\pi $ as the Gibbs
free energy. Due to our definition of ${\cal M}$, our last term is
given in terms of $B$ and not in terms of $H$. These expressions are
useful for obtaining an expression for the stress tensor in a
medium. We would like to note that there is no a unique expression
accepted by all authors \cite{Jackson}  for it in terms of the
fields $B$ and $H$. We want to state here the relation we have
obtained among the energy-momentum tensors involved in our previous
formulae, which is the sum of the Maxwell tensor for the field $B$
plus the Minkowski tensor for a nonlinear media, involving both $B$
and $H$ . As we have also $-\Omega -B{\cal M=}-\frac 1 {8\pi
}B^2+\frac{ 1}{4\pi }\int B dH+p_0$, we observe then that one can
write ${\cal T}_{ij}$ as

\begin{equation}
{\cal T}_{ij}=  p_0 \delta_{ij}+{\cal S}(B)_{i j}-{\cal T}^M_{i j}
(B, H), \label{MMIN}
\end{equation}

where ${\noindent \cal S}(B)_{i j}=\frac{1}{4\pi}[B_i B_j
-\frac{1}{2}(B^2)\delta_{i j}]$ is the Maxwell stress tensor for
the microscopic field $B$, and ${\cal T}^M_{i j}
(B)=\frac{1}{4\pi}[H_i B_j -(\int B dH)\delta_{i j}]$ is the
Minkowski tensor for nonlinear media, which reduces to the usual
expression \cite{Jackson} in the linear case. The equation
(\ref{MMIN}) expresses the way the stress tensors appear in our
problem.

If $H=0$, then $B=4\pi{\cal M}$ and ${\cal T}_{ij}= p_0 \delta_{ij}+
{\cal S}(B)_{i j}$. This gives the total pressure when the magnetic
field is maintained self-consistently, for instance, in a
ferromagnetic body.

This case deserves some comments. For the system below the
ferromagnetic phase transition (where it is yet paramagnetic), we
start from the thermodynamic potential density,   $\Omega_V= -P_0-
\int {\cal{M}} dB$. We assume that it must be an even function of
$B$, and conclude that $\cal{M}$ must be an odd function of $B$,
which we write as ${\cal{M}}=B/4\pi - a B^3+ ...$. We wrote the
second term as negative, since we consider the system below the
condition of self-magnetization. The next powers of $B$ are
neglected since they are expected to be small.

The extremum of $\Omega_V$ with regard to $B$ is determined  by the
term $\int {\cal{M}} dB$. We have $d\Omega/dB=-B/4\pi+4 a B^3=0$
which for $B
> 0$ leads to $B^2=1/(4 a \pi)$. This solution leads to a minimum
for $\Omega_V=\Omega_V (B)$, and for it $ {\cal{M}}= B/4\pi$: the
minimum of $\Omega_V$ corresponds to a self-magnetization for which
$\Omega_V=-P_0 -B^2/8\pi$. Thus  the ferromagnetic phase transition
leads to a  spontaneous symmetry breaking, leading $\Omega_V$ to
reach a minimum value for nonzero $B$. (for classical and
diamagnetic systems, as ${\cal{M}}<0$, the minimum is achieved for
$B=0$). Obviously, $a$ is a function of temperature and the
condition of self-magnetization defines the Curie temperature $T_c$.
The present discussion is especially relevant to vector Boson
systems having a magnetic moment, which is the case discussed in
refs. \cite{ChJP} and \cite{IJMPD}, in which self-magnetization
occurs.

\section{Acknowledgement}The authors acknowledge thanks the Office of
External Activities of ICTP for its support through NET-35. H.J.M.C.
is a fellow of the \textit{Funda\c{c}\~{a}o de Amparo \`{a} Pesquisa
do Estado do Rio de Janeiro} (FAPERJ), Brazil, under the contract
E-26/151.684/2002. A.P.M. thanks to TWAS and CBPF-ICRA-Br for their
hospitality and support.

\end{document}